\documentclass[amsmath,amssymb,twocolumn,aps,prd]{revtex4-1}

\usepackage{graphicx}
\usepackage{subfigure}
\usepackage{epstopdf}
\usepackage{nonfloat}
\usepackage{lineno}
\usepackage{natbib}
\usepackage{url}

\begin{document}	

\title{Mass hierarchy sensitivity of medium baseline reactor neutrino experiments with multiple detectors}

\author{Hongxin Wang$^1$}
\email{wanghongxin@ihep.ac.cn}

\author{Liang Zhan$^2$}
\email{zhanl@ihep.ac.cn}

\author{Yu-Feng Li$^2$}
\email{liyufeng@ihep.ac.cn}

\author{Guofu Cao$^2$}
\email{caogf@ihep.ac.cn}

\author{Shenjian Chen$^1$}

\affiliation{
	$^1$Department of Physics, Nanjing University, Nanjing 210093, China \\
	$^2$Institute of High Energy Physics, Chinese Academy of Sciences, Beijing 100049, China}


\begin{abstract}
We report the neutrino mass hierarchy (MH) sensitivity of medium baseline reactor neutrino experiments with multiple detectors.
Sensitivity of determining the MH can be significantly improved by adding a near detector and combining both the near and far detectors.
The size of the sensitivity improvement is related to accuracy of the individual mass-splitting measurements 
and requires strict control on the relative energy scale uncertainty of the near and far detectors.
We study the impact of both baseline and target mass of the near detector on the combined sensitivity.
A figure-of-merit is defined to optimize the baseline and target mass of the near detector and the optimal selections are $\sim$13~km and $\sim$4~kton respectively
for a far detector with the 20~kton target mass and 52.5~km baseline.
As typical examples of future medium baseline reactor neutrino experiments,
the optimal location and target mass of the near detector are selected for JUNO and RENO-50.
Finally, we discuss distinct effects of the neutrino spectrum uncertainty for setups of a single detector and double detectors, which indicate that
the spectrum uncertainty can be well constrained in the presence of the near detector.
\end{abstract}

\maketitle

\section{Introduction}

It is reported that the medium baseline reactor neutrino experiment can determine the type of the neutrino mass hierarchy (MH)
by precisely measuring the fine structure of the neutrino energy spectrum from reactors~\cite{Petcov:2001,Choubey:2003,Learned:2006wy,Zhan:2008,Zhan:2009}.
Reactor and accelerator neutrino experiments measured an unexpectedly large value of neutrino mixing angle $\theta_{13}$ in 2012~\cite{Dayabay2012,Abe:2011fz,Ahn:2012nd,Abe:2011sj,Adamson:2011qu},
which implies that the MH determination is feasible in the next one or two decades with next generation neutrino oscillation experiments.
Experiments using accelerator neutrinos with a long baseline of $\sim$1000~km~\cite{Acciarri:2015uup}, atmosphere neutrinos sensitive to the energy range of 1-20~GeV~\cite{Aartsen:2014oha,Ahmed:2015jtv},
and reactor neutrinos at a medium baseline $\sim$50~km are proposed to determine the neutrino MH~\cite{Qian:2015waa,Kettell:2013eos,Ge:2012wj,Qian:2012xh}. Among the above possibilities,
medium baseline reactor neutrino experiments, such as JUNO (Jiangmen Underground Neutrino Observatory)~\cite{Li:2013,Djurcic:2015vqa,An:2015jdp}
and RENO-50~\cite{Park:2014sja,Kim:2014rfa,Pac:2015aba,Seo:2015TAUP}, have the potential to determine the neutrino MH by using large liquid scintillator detectors ($\sim$20~kton) with
energy resolution of unprecedented levels.

Key requirements for the MH determination in reactor neutrino experiments are powerful nuclear power plants (NPPs),
large detector mass and good energy resolution.
Sensitivity study at JUNO shows that a 20~kton detector with energy resolution of $3\%/\sqrt{E_{vis}(\mathrm{MeV})}$
is mandatory to achieve a significance of better than $3\sigma$ after 6 years of running~\cite{Li:2013}.
Several interesting ideas are proposed to improve the MH sensitivity of reactor neutrino experiments,
including combining the mass splitting measurement from accelerator neutrino experiments~\cite{Li:2013},
synergy of different MH probes in reactor and atmospheric neutrino oscillation experiments~\cite{Blennow:2013}
and using two identical half-size detectors at near and far sites~\cite{Ciuffoli2013,Ciuffoli2014}.

In this work we shall discuss the MH sensitivity improvement by using the near detector (ND) and far detector (FD) in medium baseline reactor neutrino experiments.
A figure-of-merit considering both the sensitivity and experimental cost is defined to optimize the baseline and target mass of the ND.
For a fixed total mass of the ND and FD, the distribution of target mass between the ND and FD
and the baseline for the ND can be optimized. The optimization is also applied to the realistic reactor core distributions of JUNO and RENO-50.
Finally we discuss distinct effects of the neutrino spectrum uncertainty for the setups of a single detector and double detectors.

The remaining parts of this work are organized as follows. In Sec.~2 we first introduce the analysis method for the MH sensitivity in
medium baseline reactor neutrino experiments. Sec.~3 is devoted to the sensitivity improvement in the presence of the ND,
and Sec.~4 is to optimize the baseline and target mass of the ND.
Finally we discuss the impact of the energy spectrum shape uncertainty in Sec.~5 and then conclude in Sec.~6.

\section{Analysis Method}

Before considering the MH sensitivity improvement in the presence of the ND, we first calculate the sensitivity for a single detector at medium baseline.
We adopt the similar experimental parameters as those of JUNO, such as a liquid scintillator detector of 20~kton, baseline of 52.5~km, reactor thermal power of 36~GW$_{\mathrm{th}}$, energy resolution of $3\%/\sqrt{E_{vis}(\mathrm{MeV})}$ and six years of data taking.
On the other hand, the default neutrino oscillation parameters are taken as $\Delta m_{21}^2=(7.53\pm0.18)\times10^{-5}\mathrm{eV}^{2}$, $\Delta m^2 = (\Delta m_{31}^2+\Delta m_{32}^2)/2=2.48\times10^{-3}\mathrm{eV}^{2}$, $\mathrm{sin}^{2}2\theta_{13} = (9.3\pm0.8)\times10^{-2}$, and $\mathrm{sin}^22\theta_{12}= 0.846\pm0.021$~\cite{PDG2014}.
A parameterized reactor neutrino flux model in Ref.~\cite{Vogel:1989iv} is used to predict the neutrino energy spectrum for the inverse beta decay reactions in the detector.
To fully explore the fine structure of the neutrino spectrum, the spectrum was divided into 200 equal-size bins between 1.8\,MeV and 8.0\,MeV.
The neutrino event rate at the detector is calculated to be $\sim$60/day
after assuming a detection efficiency of 80\%, which is consistent with the number of JUNO in Ref.~\cite{Li:2013}.

The least squares method is used in the neutrino energy spectrum fitting and a standard $\chi^2$ function with proper
nuisance parameters and penalty terms is constructed as follows:
\begin{widetext}
	\begin{equation}
	\label{eq:Chi2Function}
	\chi^2=\sum_d\sum_{i=1}^{N_{\mathrm{bin}}}\frac{[M^d_i-T^d_i(1+\epsilon_R+\sum_r w_r\epsilon_r+\epsilon_d+\epsilon_i)]^2}
	{M^d_i}+\frac{\epsilon_R^2}{\sigma_R^2}+\sum_r\frac{\epsilon_r^2}{\sigma_r^2}+	 \sum_d\frac{\epsilon_d^2}{\sigma_d^2}+\sum_{i=1}^{N_{\mathrm{bin}}}\frac{\epsilon_i^2}{\sigma_s^2},
	\end{equation}
\end{widetext}
\noindent where \emph{d} is the detector index, \emph{i} denotes the bin number, $M$ is the measured spectrum, $T$ is the predicted spectrum,
$\epsilon$'s with different indexes are the nuisance parameters corresponding to different systematic uncertainties, $\sigma$'s with different indexes are the standard deviations
of nuisance parameters assuming the systematic uncertainty follows the Gaussian form~\cite{Stump2001}.
The systematic uncertainties include the correlated (absolute) reactor uncertainty ($\sigma_R$ = 2\%), the uncorrelated (relative) reactor uncertainty ($\sigma_r$ = 0.8\%), the spectrum shape uncertainty ($\sigma_s$ = 1\%), and the detector-related uncertainty ($\sigma_d$ = 1\%).

Variations of $\sin^22\theta_{12}$ and $\Delta m^2_{21}$ within their allowed ranges have negligible effects on the best-fit $\chi^2$ value. 
Precise measurement of $\sin^22\theta_{13}$ with a $3\%$ uncertainty is expected from Daya Bay experiment after 2017, and the variation induced to the best-fit $\chi^2$ value is $\sim 0.2$.
For fast minimization of $\chi^2$ function, we fixed all the mixing angles and mass splitting except for $\Delta m^2$.
Thus, the $\chi^2$ value is a function of $\Delta m^2$.
The best fit value $\chi^2_{\mathrm{min}}$ can be obtained through scanning $\Delta m^2$.

The discriminator of MH can be obtained using both the normal hierarchy (NH) and inverted hierarchy (IH) models to fit the simulation neutrino spectrum generated by the NH model:
\begin{equation}
\label{eq:DeltaChi2}
\Delta\chi^2=|\chi^2_{\mathrm{min}}(\mathrm{NH})-\chi^2_{\mathrm{min}}(\mathrm{IH})|.
\end{equation}
The simulation studies are also accomplished by assuming the IH model as the true one, which gives the consistent conclusion with the assumption of the NH model.
In the following, we shall illustrate the simulation results of the NH model.

When we apply the calculation for the default case of a single detector as JUNO,
the MH sensitivity is found to be $\Delta \chi^2 \sim 16.3$, and consistent with the results in Ref.~\cite{Li:2013} without considering the real reactor core distribution.

\section{Sensitivity improvement due to near detector}

Now we want to add a ND and calculate the combined MH sensitivity by adding the neutrino energy spectrum information of the ND in Eq.~(\ref{eq:Chi2Function}).
A common set of oscillation parameters is used in the spectrum predictions of the ND and FD. 

\begin{figure}[tbph!]
\begin{center}		
\includegraphics[width=\columnwidth]{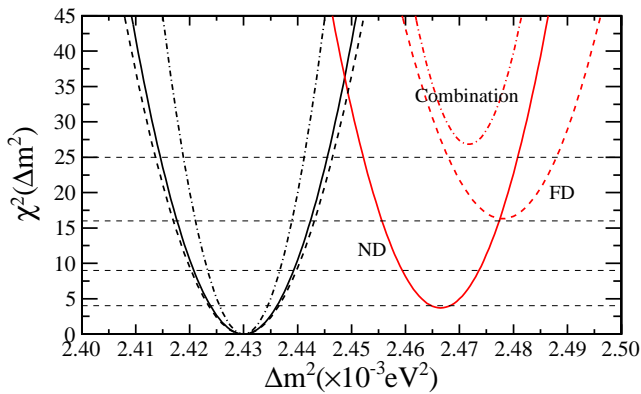}	
\figcaption{Three classes of $\chi^2$ values as functions of $\Delta m^2$ for the ND (solid lines), FD (dash lines) and the combined ND and FD (dot-dash lines).
The true NH is assumed and the best-fit values of the IH case (red lines) are the MH sensitivity defined in Eq.~\ref{eq:Chi2Function}.}
\label{fig:principle}
\end{center}
\end{figure}

To illustrate the improvement of the MH sensitivity, a ND with the target mass of 10~kton and baseline of 30~km is assumed as an initial choice.
The dot-dashed, solid and dashed lines in Fig.~\ref{fig:principle} show the $\chi^2$ as the functions of $\Delta m^2$.
The true NH is assumed to generate the experiment energy spectrum and the black lines is the $\chi^2$ value using the NH model to fit the energy spectrum.
Therefore, the best fit (minimal) values of $\chi^2$ for the NH case are 0. On the other hand, the best fit values of $\chi^2$ for the IH case equal
to the MH discriminator defined in Eq.~\ref{eq:DeltaChi2}, which are 26.8, 3.7 and 16.3 respectively. 
Comparing different combinations of the ND and FD, we have $\chi^2_{\mathrm{com, min}} > \chi^2_{1,\mathrm{min}} + \chi^2_{2,\mathrm{min}}$, which demonstrates the improvement of the MH sensitivity by combining the near and far detectors, where $\chi^2_{\mathrm{com, min}}$, $\chi^2_{1,\mathrm{min}}$ and $\chi^2_{2,\mathrm{min}}$ are for the sensitivity of combined, ND, and FD scenarios.

The improvement of the combination of the ND and FD can be explained as follows.
Using the standard least squares method, the $\chi^2$ distribution is approximatively parabolic and can be expressed as
\begin{equation}
\chi^2 \simeq \chi_{\mathrm{min}}^2 + \left(\frac{x-x_{\mathrm{best}}}{\sigma}\right)^2,
\end{equation}
where \emph{x} denotes the variable $\Delta m^2$, $x_\mathrm{best}$ and $\sigma$ denote the best-fit and uncertainty of $\Delta m^2$. $\chi_{\mathrm{min}}^2$ is the best fit (minimal) value of $\chi^2$.
We use $\chi^2_i$ (i=1,2) to represent the $\chi^2$ function for the ND and FD respectively.
The combined $\chi^2$ can be approximated as $\chi^2_{\mathrm{com}} \simeq \chi^2_1 + \chi^2_2$ when the statistical uncertainty dominates.
The combined best-fit value of $\Delta m^2$ can be expressed analytically as
\begin{equation}
\label{eq:BestComX}
x_{\mathrm{com,best}} \simeq \frac{x_{\mathrm{1,best}}\sigma^2_2+x_{\mathrm{2,best}}\sigma_1^2}{\sigma_1^2+\sigma_2^2}.
\end{equation}
Hence the corresponding best-fit value of the combined $\chi^2$ is
\begin{eqnarray}
\label{eq:ComChi2}
\chi^2_{\mathrm{com}} = \chi^2_{\mathrm{1,min}} + \chi^2_{\mathrm{2, min}} + \chi^2_{\mathrm{ext}},  \\ \nonumber
\chi^2_{\mathrm{ext}} = \frac{(x_{\mathrm{1,best}}-x_{\mathrm{2,best}})^2}{\sigma_1^2+\sigma_2^2},
\end{eqnarray}
where $\chi^2_{\mathrm{1,min}}$ ($\chi^2_{\mathrm{2,min}}$) is the respective sensitivity of the MH determination for the ND (FD), and $\chi^2_{\mathrm{ext}}$ is
the extra MH sensitivity because of the combination.
$\chi^2_{\mathrm{ext}}>0$ due to different best-fit values of $\Delta m^2$, as shown in Fig.~\ref{fig:principle}.
Obviously, the extra MH sensitivity is related to the difference of the $\Delta m^2$ best-fit values and their uncertainties in the ND and FD.
Sensitivity increase can be obtained even when there is no or very poor MH discrimination ability in the ND,
as long as the ND can provide a different best-fit value of $\Delta m^2$.
\begin{figure}[tbph!]
	\begin{center}
		\includegraphics[width=\columnwidth]{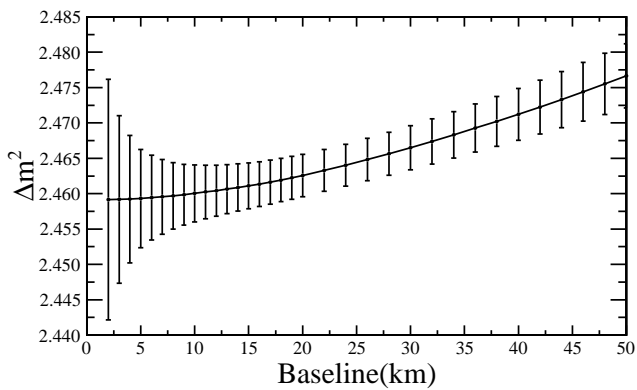}		
\figcaption{The best-fit value and uncertainty of $\Delta m^2$ as a function of baseline for the ND.
The true MH is the NH model and the fitted MH is the IH model.}
		\label{fig:DmError}
	\end{center}
\end{figure}

\begin{figure}[tbph!]
	\begin{center}	
	\includegraphics[width=\columnwidth]{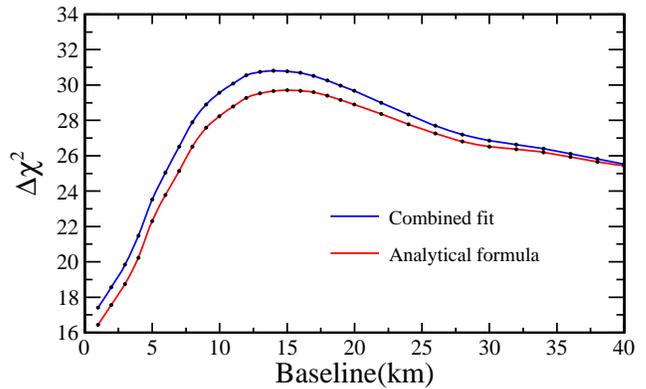}
	\figcaption{Combined MH sensitivity of the ND and FD as a function of the near detector baseline.
    The blue and red lines are the results of calculation using the combined fit and analytical formula.}
	\label{fig:validation}
	\end{center}
\end{figure}

The best-fit value and uncertainty of $\Delta m^2$ as a function of baseline for the ND is shown in Fig.~\ref{fig:DmError},
which are used as the inputs of Eq.~\ref{eq:ComChi2} to calculate the combined sensitivity $\chi^2_{\mathrm{com}}$.
On the other hand, we can also directly calculate the combined sensitivity by fitting Eq.~\ref{eq:Chi2Function} with the inputs of both the ND and FD.
As shown in Fig.~\ref{fig:validation}, these two results are rather consistent, and thus Eq.~\ref{eq:ComChi2} is a good approximation of the combined sensitivity.
The optimal baseline to maximize the MH sensitivity for the ND of 10~kton is around 15~km.

In our current study, we have neglected the possible detector systematics uncertainties. 
As one can see from Eq.~\ref{eq:ComChi2}, the sensitivity improvement due to $\chi^2_{\mathrm{ext}}$ 
depends on the uncertainties of the mass-splitting measurements in both the ND and FD. 
In our case, $\sigma_1$ and $\sigma_2$ are dominated by the statistical uncertainties. 
The relative difference between two best-fit values of the mass-splitting measurements (i.e., $x_{\mathrm{1,best}}-x_{\mathrm{2,best}}$) is about 0.7\%. As a result, an uncertainty of the mass-splitting at the level of 1\% will largely reduce $\chi^2_{\mathrm{ext}}$.
However, only the uncorrelated uncertainties in the mass-splitting measurements contribute to the best-fit difference and the correlated uncertainties will be cancelled out. 
The main systematic uncertainties for the mass-splitting measurement are the energy scale uncertainties. 
The MH sensitivity improvement requires strict control of relative energy scale uncertainties, e.g., below the level of 0.5\%.

\section{Optimization of target mass and baseline}

For the experiment of a single detector like JUNO~\cite{Li:2013}, the baseline was optimized at $\sim$52.5~km for the MH determination.
The sensitivity $\Delta \chi^2$ is approximately proportional to the target mass.
However the target mass is constrained by the technical challenges and experiment cost, and current selection of the target mass is 20~kton.
Using a ND as the combination with the FD, the requirement for the FD target mass can be reduced.
A proper selection of target mass and baseline of the ND has the possibility to improve the MH sensitivity even the total target mass of the ND and FD keeps as 20~kton.

In the optimization of the target mass and baseline for the ND,
we first fixed the target mass and baseline for the far detector,
and therefore $\chi^2_{\mathrm{2,min}}$, $x_{\mathrm{2,best}}$, $\sigma_2$ are also fixed in Eq.~\ref{eq:ComChi2}.
Given a target mass of the ND, $x_{\mathrm{1,best}}$ and $\Delta \chi^2_{\mathrm{ext}}$ are functions of baseline, and an optimal baseline can be obtained by maximizing the combined sensitivity $\Delta \chi^2_\mathrm{{com}}$.
At the optimal baseline of near detector, a larger target mass will always increase the MH sensitivity because of the larger statistics from the ND.
However, larger target mass will increase the cost and technical challenges of the experiment construction. In this respect,
we propose a figure-of-merit defined as
\begin{equation}
\label{eq:FigureMerit}
F = \frac{\Delta\chi^2_{\mathrm{com}}}{M_1 + M_2},
\end{equation}
where $M_1$ and $M_2$ are the target mass of the near and far detectors respectively, and
$F$ denotes the optimal sensitivity per target mass.
Given a total target mass, the target mass ratio between the ND and FD can be optimized by maximizing $F$.
The current proposals of the JUNO and RENO-50 experiments are special cases of $M_1 = 0$, and $F$ is almost constant
because $\Delta\chi^2$ is approximately proportional to the target mass because the statistics dominates.

In the following, we shall first study the optimization of the target mass and baseline in the ideal case
where the real reactor core distribution is not taken into account,
and then study the optimization for the realistic reactor core distributions of JUNO and RENO-50.
The optimal ND location and target mass for JUNO and RENO-50 are provided.

\subsection{Ideal case}
First we consider the ideal case with a single baseline from the reactor to the detector.
Similar to JUNO, we assume the total reactor thermal power to be 36~GW, and the baseline of the FD is fixed at 52.5~km.
For the configuration of a single detector, the figure-of-merit is $F\simeq16.3/20\simeq0.815$,
which is almost constant when the detector target mass varies.

\begin{figure}[htbp!]
	\centering
	\subfigure[M$_2$=10 kton]{\label{fig:a}\includegraphics[width=0.9\columnwidth]{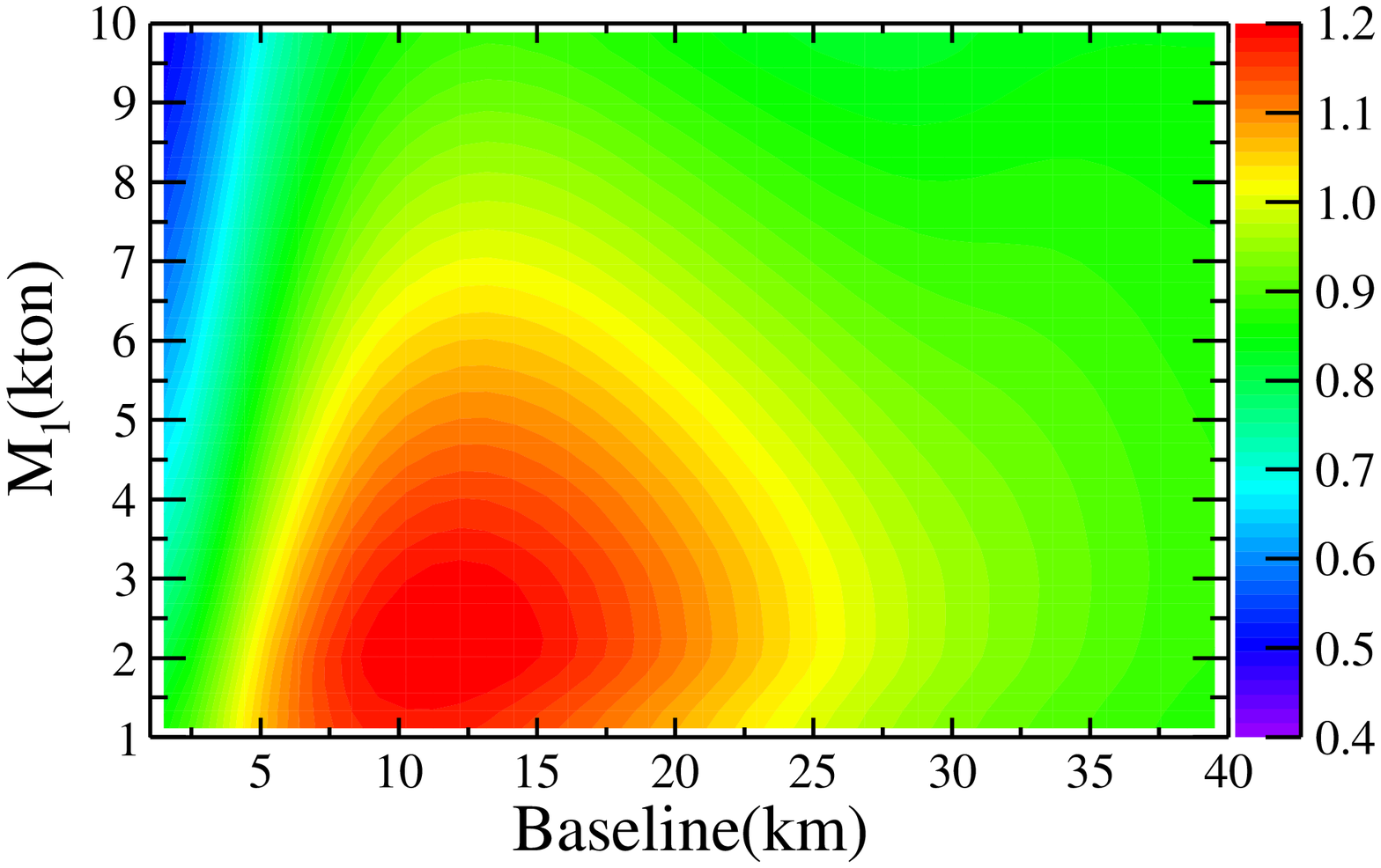}}
	\subfigure[M$_2$=20 kton]{\label{fig:b}\includegraphics[width=0.9\columnwidth]{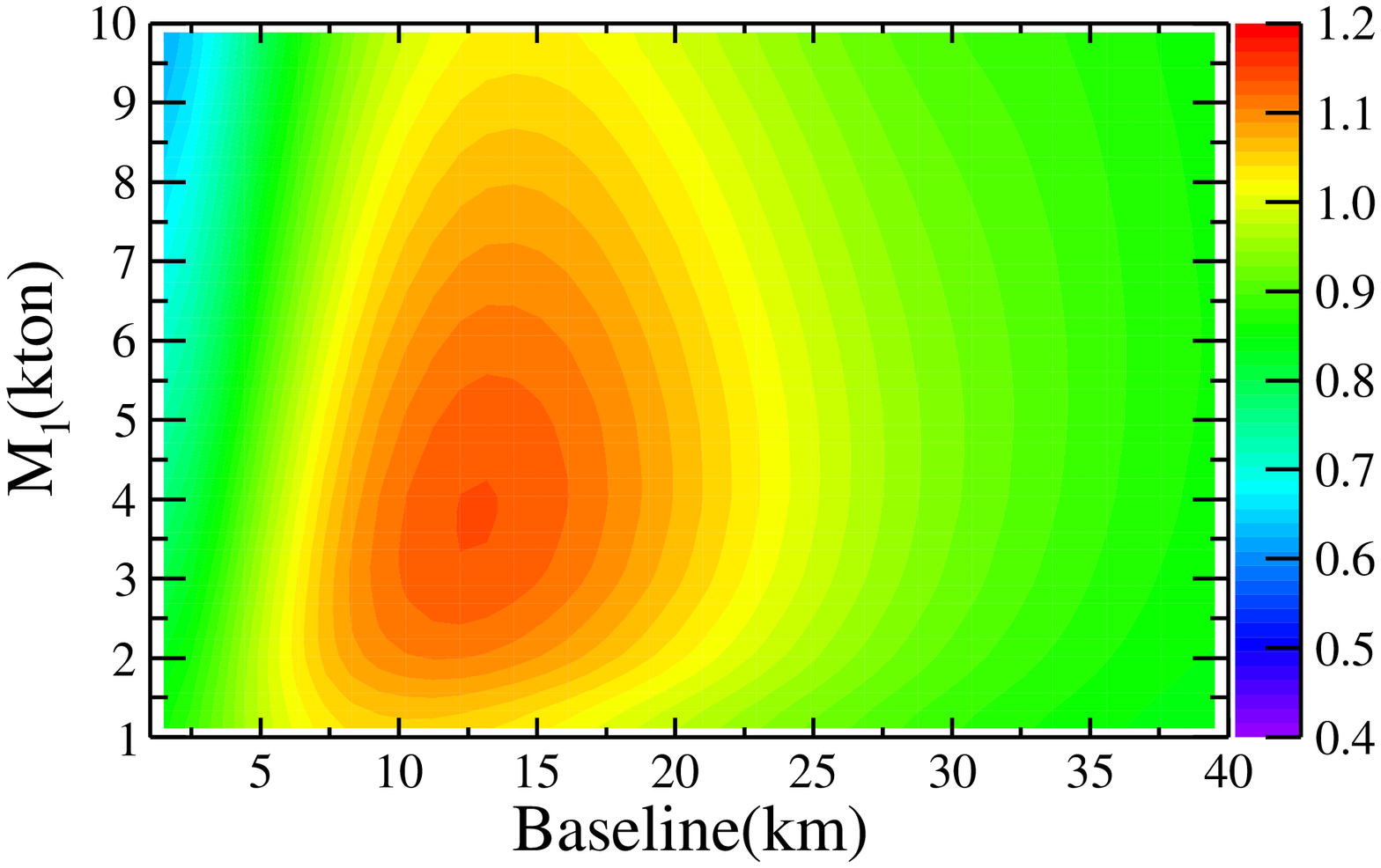}}
	\subfigure[M$_2$=30 kton]{\label{fig:c}\includegraphics[width=0.9\columnwidth]{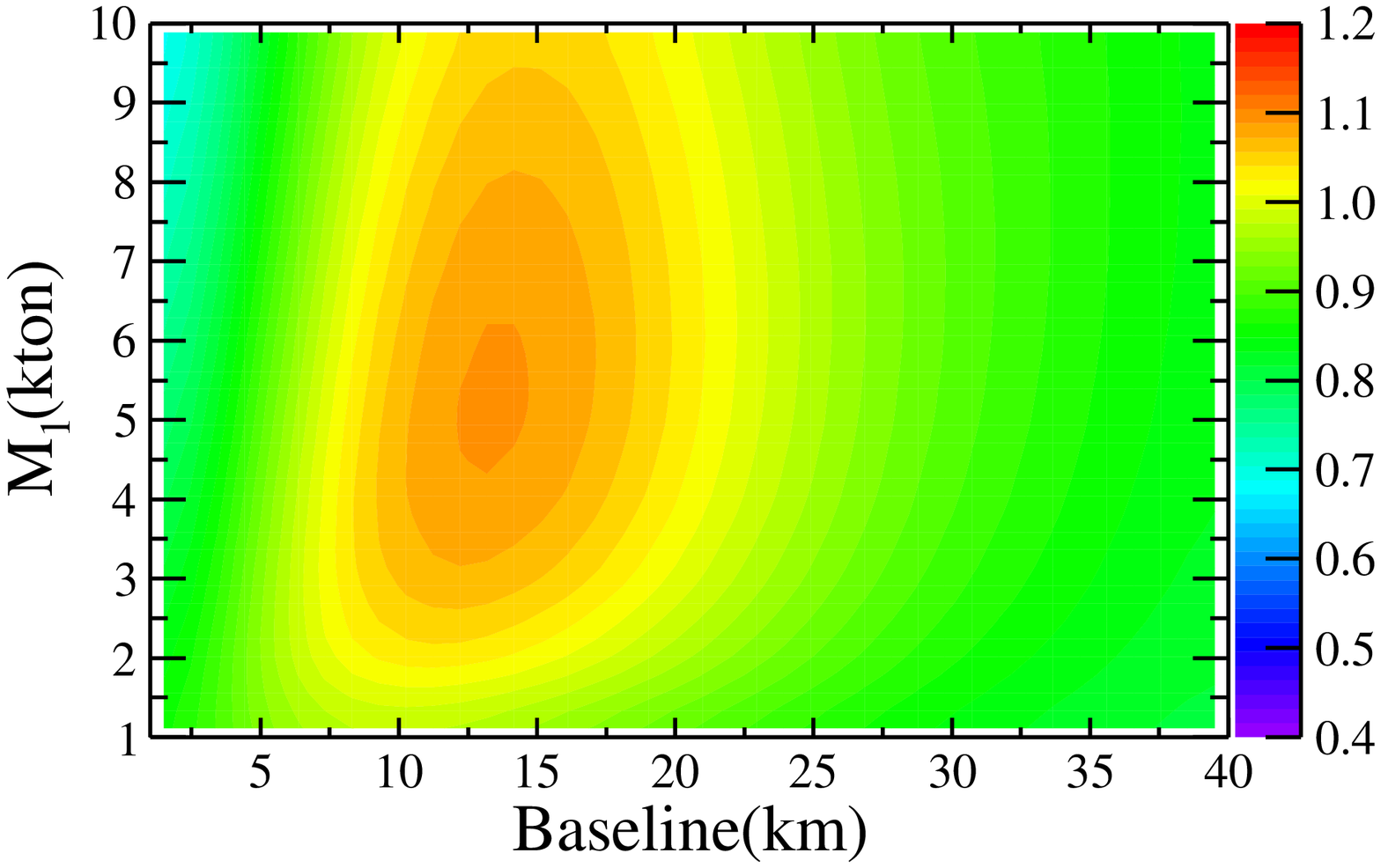}}
	\subfigure[M$_2$=40 kton]{\label{fig:d}\includegraphics[width=0.9\columnwidth]{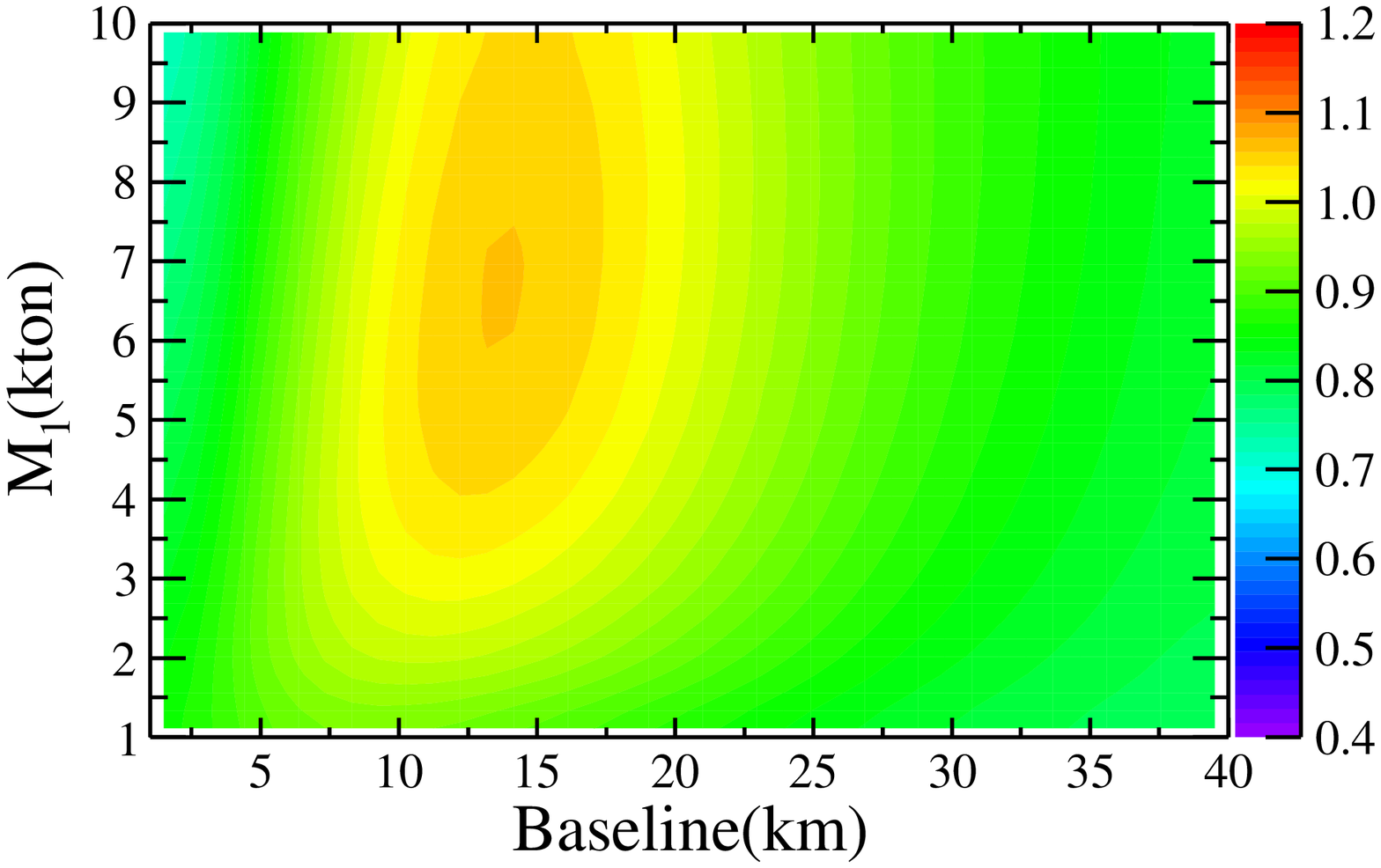}}
	\caption{The contours of the figure-of-merit $F$, as a function of target mass and baseline for the ND when fixing the target mass of the FD at several typical values.
A maximal value of $F$ indicates the optimal target mass and baseline for the ND to maximize the MH sensitivity per total target mass.}
	\label{fig:2DforDiffM2}
\end{figure}

The target mass of the FD was set to be several typical values as 10~kton, 20~kton, 30~kton and 40~kton.
Given a target mass of the FD, we change the target mass and baseline of the ND and calculate the figure-of-merit $F$ for optimization.
The results are shown in Fig.~\ref{fig:2DforDiffM2}, where in a large parameter space of the ND target mass and baseline,
the MH sensitivity can be improved in comparison to the single detector configuration. 
When the baseline is too small, the double detector coulb be worse than the single detector configuration,
because $\chi^2_{\mathrm{1,min}}\simeq0$ and $\chi^2_{\mathrm{ext,min}}$ can not compensate the
contribution of adding the same target mass at the FD.
The optimal target mass is $\sim$4~kton and the baseline is $\sim$13~km for a 20~kton FD.
The contours of the figure-of-merit in Fig.~\ref{fig:2DforDiffM2} show that the optimal baseline is
in the region of 10-15~km, and is approximately independent on the target mass of the FD.
However, the optimal target mass of the ND depends on the FD target mass.

\begin{figure}[htbp!]
	\begin{center}
		\includegraphics[width=\columnwidth]{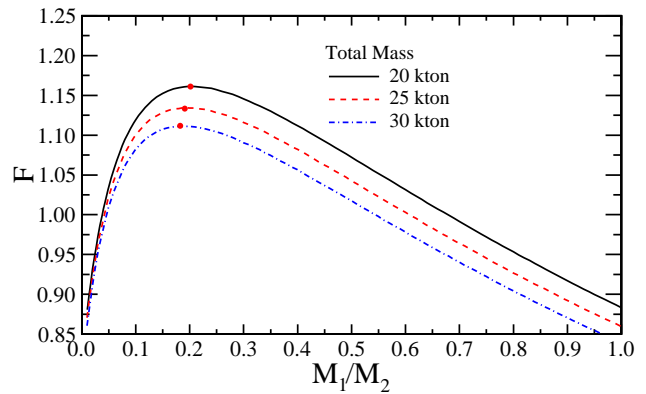}
		\figcaption{Figure-of-merit $F$ as function of the target mass ratio of the near detector to the far detector at different total target mass.
As a comparison, we have $F=0.815$ for a single 20-kton far detector.}	
		\label{fig:chooseRatio}	
	\end{center}
\end{figure}

We then fix the baseline of the ND to be 13~km, and vary the target mass ratio between the ND and FD for different total target masses.
The result is shown in Fig.~\ref{fig:chooseRatio}, where the optimal ND target mass is changing with the FD target mass,
but the target mass ratio $M_1/M_2$ approximately keeps rather stable at $\sim$0.2.

\subsection{JUNO}

In the realistic case, there are multiple reactor cores in one NPP and the baseline from the detector to the reactor cores can not be identical.
The difference of multiple baselines will reduce the sensitivity of the MH determination as studied in~\cite{Li:2013}.
In this section, we shall study the MH sensitivity improvement due to the ND for JUNO. The selection of the baseline and the target mass of the ND are optimized.

There are ten reactor cores in Yangjiang and Taishan NPPs for the JUNO experiment.
We adopt the baseline and power setups of the reactor cores listed in \cite{Li:2013} in our MH sensitivity calculation.
We obtain $\Delta \chi^2 $ = 11.6 for JUNO with the realistic reactor core distribution,
while we have $\Delta \chi^2 $ = 16.3 for the ideal case using the identical baseline from Yangjiang and Taishan reactors.
These results are consistent with the calculation in Ref.~\cite{Li:2013}. Therefore, we have $F\simeq11.6/20\simeq0.58$ for the current JUNO far detector.
\begin{center}
	\includegraphics[width=8cm]{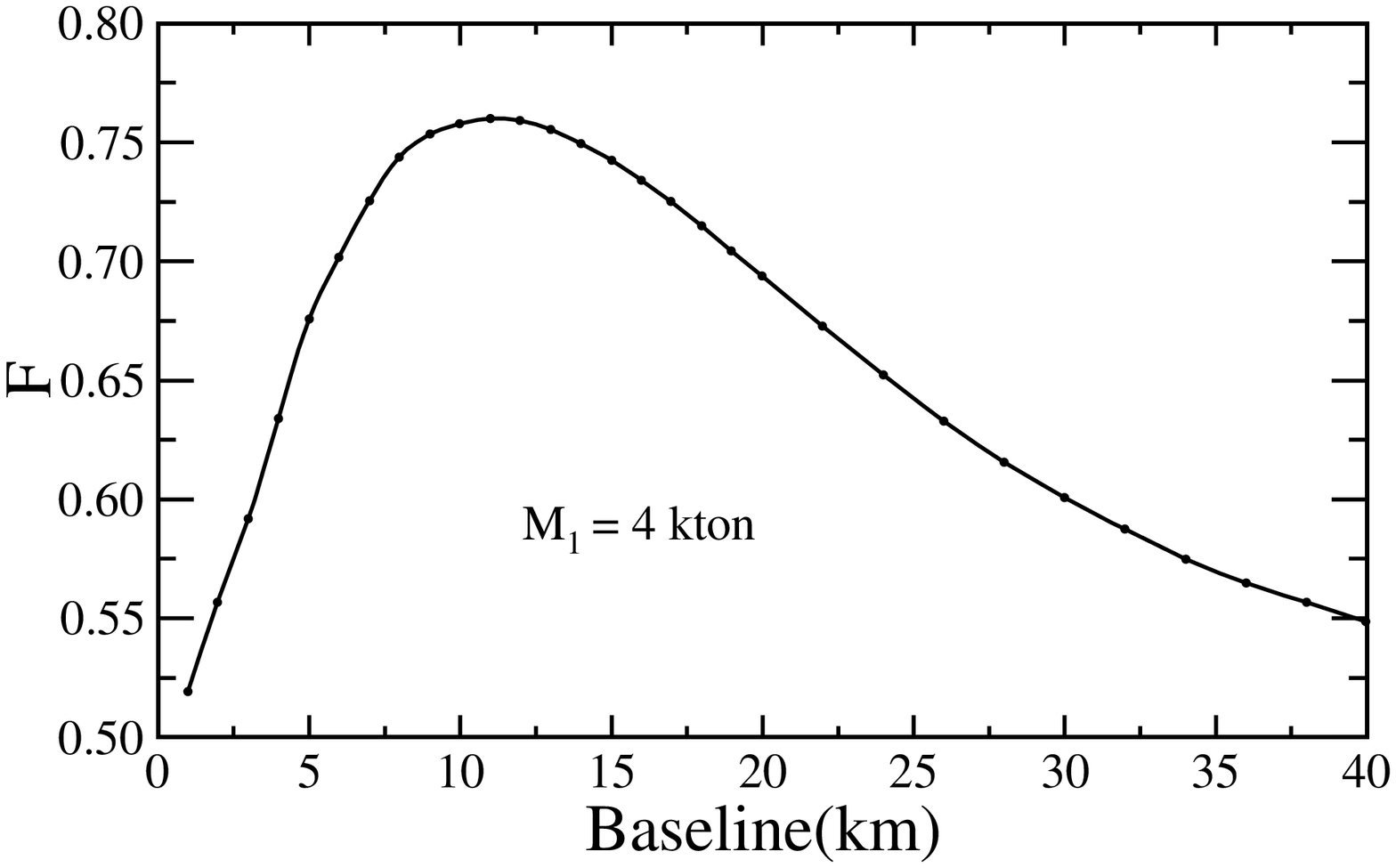}
	\figcaption{\label{fig:JUNOtwodetector}
Figure-of-merit for the combined ND and FD for JUNO as a function of the ND baseline.
As a comparison, we have $F\simeq0.58$ for the current JUNO far detector.}
\end{center}

The distance between Yangjiang and Taishan NPPs is ~77~km,
therefore, there is no proper position for one ND at the baseline of 10-20~km from both the Yangjiang and Taishan NPPs.
We consider the ND for Yangjiang and Taishan separately.
For the Yangjiang NPP, there is no good location for a ND because the possible location is in the sea in the perpendicular line of the six reactor cores.
Hence we only consider the ND for the Taishan NPP.
The locations of the planed four reactor cores in Taishan are identified using the map of Google Earth.
We intend to find a location with almost equal baselines from the four reactor cores.
The actual baseline differences from the four Taishan reactor cores to the possible candidate location are around 0.01~km.
The figure-of-merit $F$ with the combination of the current JUNO detector and the ND with a target mass of 4~kton is shown in Fig.~\ref{fig:JUNOtwodetector}.
As a comparison, the figure-of-merit is $F\simeq0.58$ for the current JUNO FD.
In Fig.~\ref{fig:JUNOtwodetector}, when the ND baseline increases, $F$ becomes smaller than 0.58 because of the interference of the Yangjiang NPP to the ND.
As shown in the following, this effect disappears for RENO-50 because there is only one NPP in the calculation.
In conclusion, a ND with the baseline of 11~km and the target mass of 4~kton can improve the MH sensitivity by $\Delta \chi^2 $ = 6.62,
while a target mass of 24~kton in the current JUNO site only improve the sensitivity by 2.32.

\subsection{RENO-50}

We can study the MH sensitivity improvement due to the ND for RENO-50.
RENO-50 plans to build a 18 kton detector located at Mt. GuemSeong with a baseline of $\sim$47~km from the Hanbit NPP of YongGwang with the total thermal power of 16.5\,GW assuming energy resolution of $3\%/\sqrt{E_{vis}(\mathrm{MeV})}$~\cite{Kim:2014rfa}.
We calculate the MH sensitivity for RENO-50 and find $\Delta\chi^2\simeq6.87$ for six years of data taking.
It implies that more than $3\sigma$ significance can be obtained from data of $\sim$10 years, and this conclusion is consistent with the calculation in Ref.~\cite{Seo:2015TAUP}.

The figure-of-merit is $F\simeq6.87/18\simeq0.38$ for RENO-50. The difference of the figure-of-merit between JUNO and RENO-50 is mainly due to the reactor power.
And then we add a ND of 4~kton and calculate $F$ as a function of the ND baseline as shown in Fig.~\ref{fig:RENOtwodetector}.
In the selection of the ND site, we have kept the baseline difference of reactor cores as small as possible.
The actual baseline difference for the candidate site is about 0.018~km.

\begin{center}
	\includegraphics[width=\columnwidth]{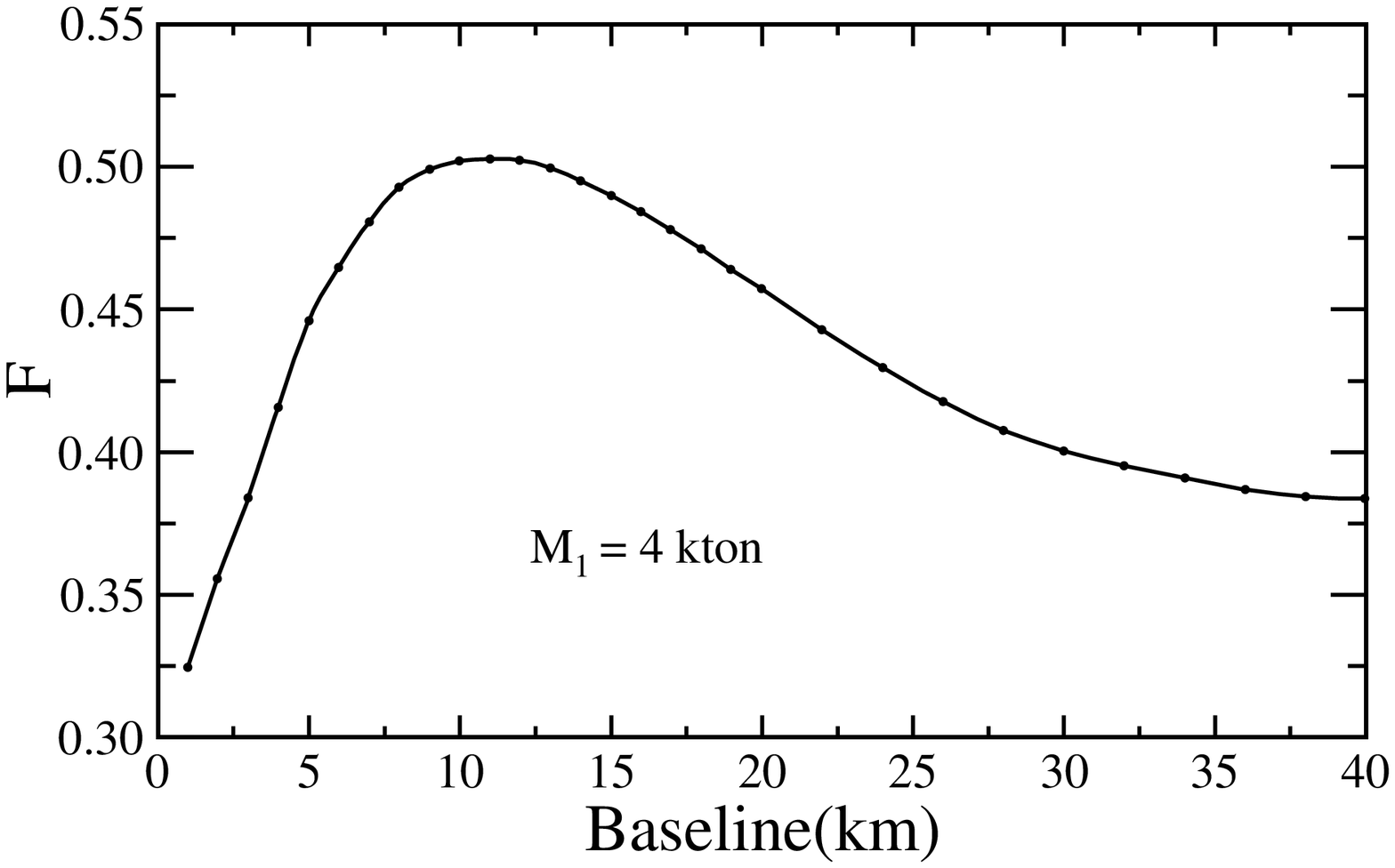}
	\figcaption{\label{fig:RENOtwodetector}
	The figure of merit for the combined ND and FD for RENO-50 as a function of the ND baseline.
As a comparison, $F=0.38$ for the current RENO-50 FD.}
\end{center}

\section{Impact of the energy spectrum shape uncertainty}

The new observed event excess at $\sim$5 MeV~\cite{An:2015nua,RENO:2015ksa,Abe:2014bwa} as well as the recent re-evaluations of the reactor neutrino flux indicate that the
reactor energy spectrum shape uncertainty could be underestimated~\cite{Hayes:2013wra,Dwyer:2014eka,Hayes:2015yka}. The reactor shape uncertainty could be 4\% or even larger.
Therefore, it is interesting to investigate the effect of the reactor shape uncertainty in the double detector configuration.

\begin{figure}[tbph!]
	\begin{center}
		\includegraphics[width=\columnwidth]{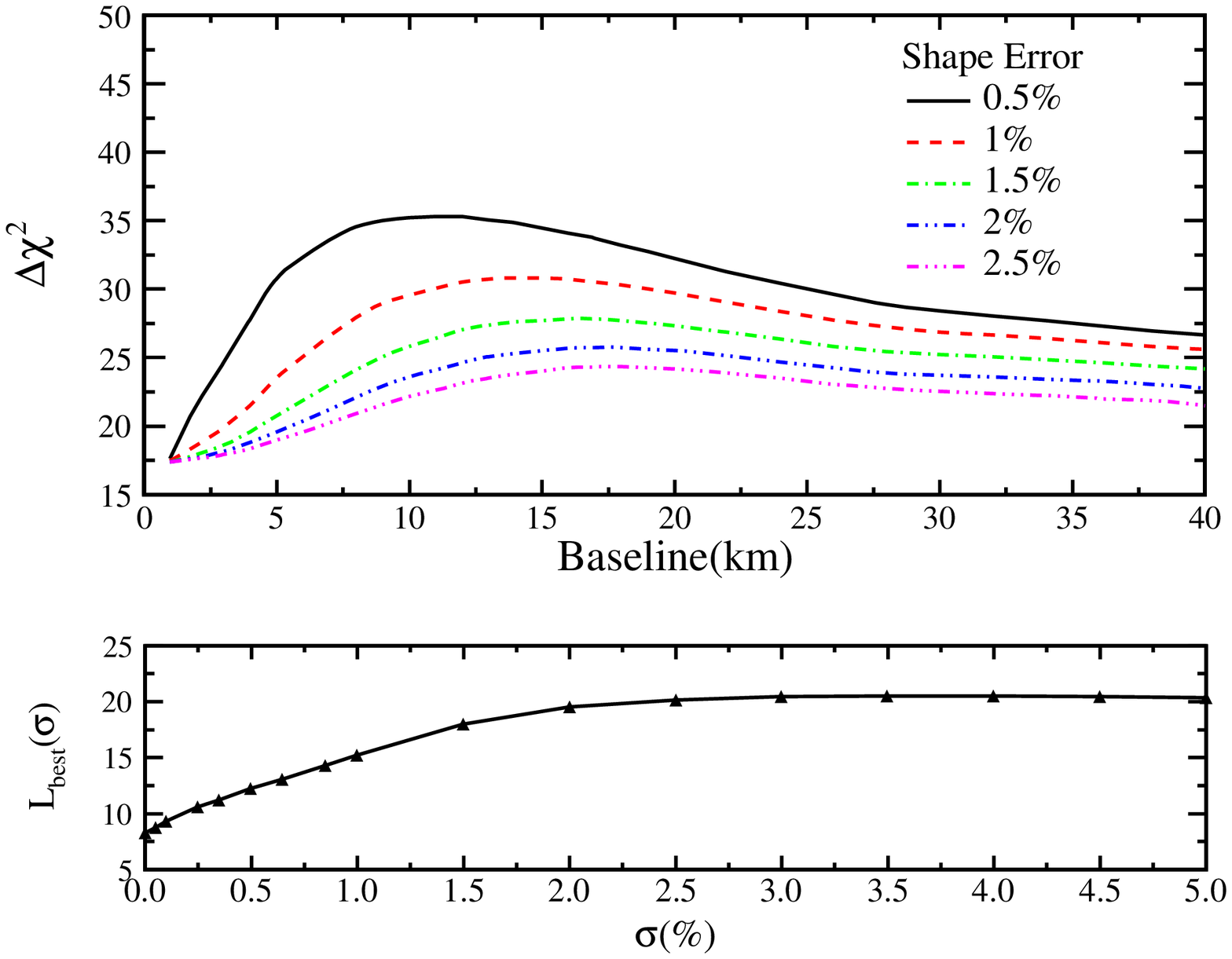}
		\figcaption{The combined MH sensitivity as a function of the ND baseline (top panel),
         and the optimal baseline of the ND as a function of the reactor shape uncertainty (bottom panel).}
		\label{fig:shapeToBaseline}
	\end{center}
\end{figure}

As shape uncertainty changes, the uncertainty of $\Delta m^2$ in the ND shown in Fig.~\ref{fig:DmError} changes accordingly,
and it will affect the optimization of the baseline for the ND.
Following the same FD setup as in Sec.~IV-A,
we fix the ND target mass as 10~kton and study the relation between the optimal ND baseline and the size of the reactor shape uncertainty,
which is showed in Fig.~\ref{fig:shapeToBaseline}.
The top panel of Fig.~\ref{fig:shapeToBaseline} shows the $\Delta\chi^2$ as a function of the baseline for different reactor shape uncertainties.
and the bottom panel of Fig.~\ref{fig:shapeToBaseline} shows the optimal baseline of of the ND as a function of the reactor shape uncertainty.
The optimal baseline varies between 8-20~km and it increases when the reactor shape uncertainty becomes larger.

\begin{figure}[tbph!]
	\centering
	\includegraphics[width=\columnwidth]{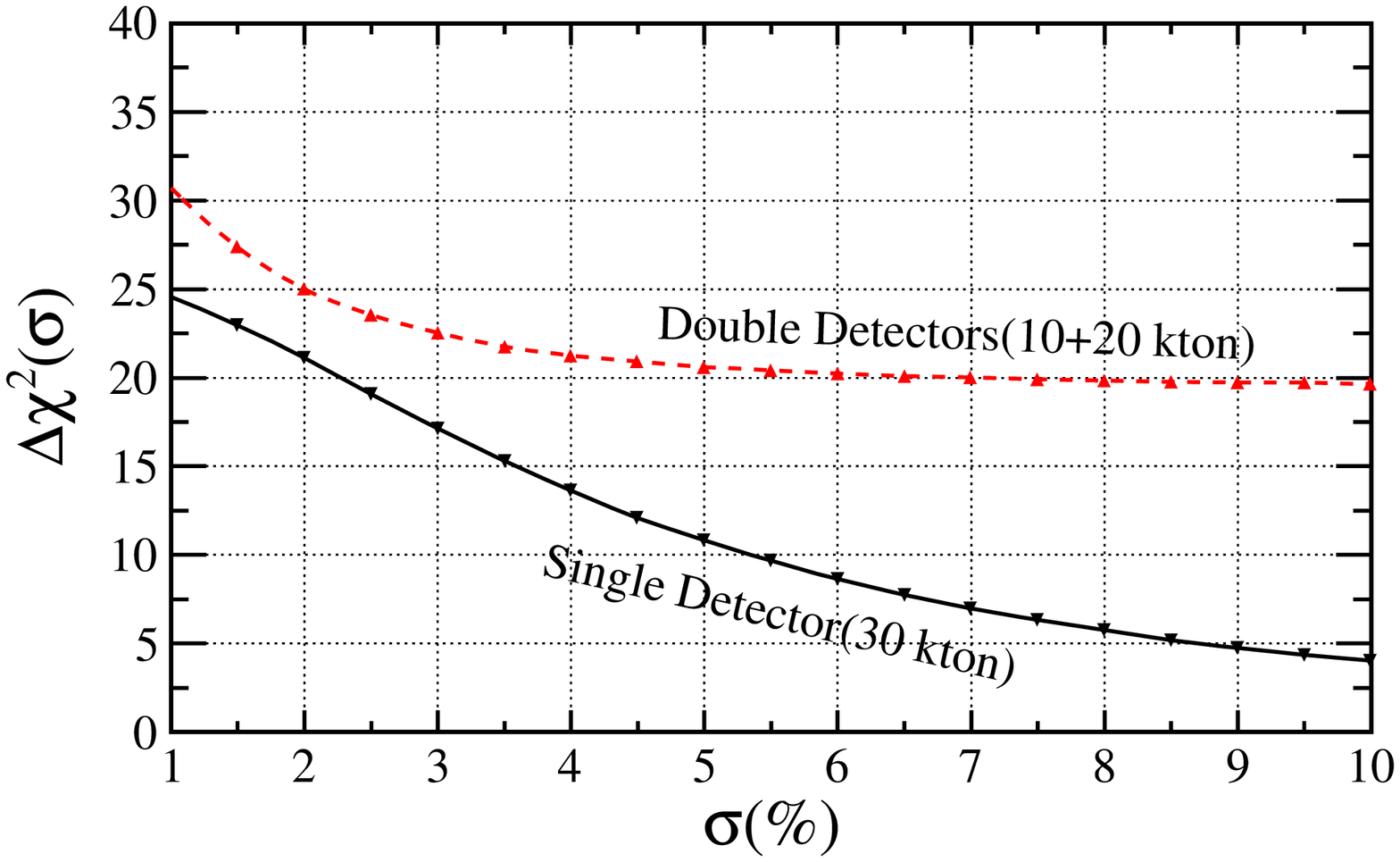}
	\figcaption{The MH sensitivity as a function of the reactor shape uncertainty for both the single (solid line) and double (dashed line) detector configurations.}
	\label{fig:constraint}
\end{figure}

Another possible way to show the MH sensitivity improvement due to the ND can be revealed by the near-far relative measurement.
In case of a larger reactor shape uncertainty for the FD, the precise reactor spectrum measurement from the ND can be used to constrain the shape uncertainty.
As a comparison, we assume a 10~kton near detector located at 13~km from the reactor and a 20~kton far detector with the baseline of 52.5~km as the double detector configuration,
and a 30~kton detector located at 52.5~km from the reactor for the single detector configuration.
The MH sensitivity as a function of the reactor shape uncertainty is showed in Fig.~\ref{fig:constraint}.
The solid line shows the sensitivity of the single detector configuration, and the dash line is the sensitivity for the double detector configuration.
With the increase of the reactor shape uncertainty, $\Delta\chi^2$ of the single detector configuration reduces rapidly, while the $\Delta\chi^2$ of
the double detector configuration first reduces and then becomes stable.
Fig.~\ref{fig:constraint} shows that the MH sensitivity of the single detector configuration with the shape uncertainty of 1\% (2\%)
approximates to that of the two detector configuration with the shape uncertainty of 2.1\% (4.2\%) for the same total target mass.
When the shape uncertainty is approaching to infinity, $\Delta\chi^2$ of the single detector configuration will be close to zero and
no information can be extracted from the energy spectrum.
However due to the constraint of the ND, the double detector configuration can determine the MH with a high sensitivity.

\section{CONCLUSION}

In this work we have studied the MH sensitivity of medium baseline reactor neutrino experiments with multiple detectors.
Sensitivity can be improved by combing the near and far detectors but requires strict control on the relative energy scale uncertainties.
A figure-of-merit is constructed to optimize the baseline and target mass of the near detector.
Results are presented in the ideal case with the identical baseline, and the realistic cases for JUNO and RENO-50.
In addition, due to the constraint of the near detector in the neutrino energy spectrum measurement, the double detector configuration
can reduce the impact of the shape uncertainty from the reactor neutrino flux prediction.

\section*{Acknowledgments}
This work was supported in part by the National Natural Science Foundation of China under Grant Nos. 11390383, 11135009, 11205176 and 11305193, by the Youth Innovation Promotion Association CAS, by the Strategic Priority Research Program of the Chinese Academy of Sciences under Grant No. XDA10010100, and the CAS Center for Excellence in Particle Physics (CCEPP).

\bibliographystyle{apsrev4-1}
\bibliography{reference}

\end{document}